# Ground-based Solar Observations for Space Weather Forecasting

A.G. Tlatov[1], A.A. Pevtsov[2]

[1]*Kislovodsk Mountain Astronomical Station of the Pulkovo Observatory, Kislovodsk, 357700, Russia*
[2]*National Solar Observatory, Boulder, CO 80303, USA*

The possibilities of organizing an observation service for solar activity in order to provide space weather forecasting are considered. The most promising at this stage is the creation of a ground-based observation network. Such a network should include solar magnetographs that provide observation of large-scale magnetic fields of the Sun, and patrol optical telescopes designed to detect coronal mass ejections and solar flares. The data of magnetographic observations provide an assessment of recurrent solar winds. Patrol telescopes operating in continuous mode allow detecting the moments of eruption and determining the parameters of coronal mass ejections at the initial stage of acceleration. The network service can be supplemented with other types of observations in the radio and optical bands. The paper considers the composition of observational tools, as well as methods and models for forecasting.

# Наземные наблюдения Солнца для прогнозирования космической погоды

А.Г. Тлатов[1], А.А. Певцов[2]

[1]*Кисловодская Горная астрономическая станция ГАО РАН*
[2]*National Solar Observatory, Boulder, Colorado 80303, USA*

Рассмотрены возможности организации наблюдательной службы за солнечной активностью в целях обеспечения прогнозирования космической погоды. Наиболее перспективным на данном этапе является создание наземной наблюдательной сети. В составе такой сети должны быть солнечные магнитографы, обеспечивающие наблюдение крупномасштабных магнитных полей Солнца, и патрульные оптические телескопы, предназначенные для детектирования корональных выбросов массы и солнечных вспышек. Данные магнитографических наблюдений обеспечивают оценку рекуррентных солнечного ветра. Патрульные телескопы, работающие в непрерывном режиме, позволят детектировать моменты эрупции и определять параметры корональных выбросов массы на начальном этапе разгона. Сетевая служба может дополняться другими видами наблюдений в радио и оптическом диапазонах. В работе рассмотрены состав средств наблюдений, а также методов и моделей для обеспечения прогнозирования.

## 1. Введение

В настоящее время одними из наиболее актуальных задач в исследовании Солнца и солнечно-земных связей является изучение и прогноз космической погоды (КП), вызываемой солнечной активностью. Различные проявления космической погоды могут влиять на многие технологические системы на Земле в воздушном пространстве и космосе. Солнечные вспышки могут производить мощное рентгеновское излучение, которое ухудшают или блокируют высокочастотные радиоволны, используемые для радиосвязи. Высокоэнергичные солнечные частицы (энергетические протоны) могут проникать в бортовую электронику спутников и приводить к сбоям космических аппаратов. Геомагнитные бури, могут изменять сигнал радионавигационных систем (ГЛОНАС, GPS,GNSS), понижая точность геопозиционирования. Геомагнитные бури также вызывают полярное сияние.

В настоящее время центральное место в составлении текущих прогнозов и прогнозировании солнечных эффектов занимает наблюдения солнечных магнитных полей, структуры короны, которая зависит от этих полей, и связанных с ними электромагнитных излучений в широком диапазоне длин волн от радио до рентгеновских лучей. Наиболее востребованной информацией для прогнозирования КП являются: 1) уровень солнечной активности в виде вспышек, высокоскоростных потоков солнечного ветра и корональных выбросов массы (КВМ), которые являются предвестниками нарушений работы таких систем, как электрические сети и трубопроводы, Глобальной системы позиционирования (ГЛОНАС, GPS) и других космических аппаратов, а также запусков космических аппаратов; 2) оценка потоков солнечного излучения в экстремальном ультрафиолетовом (EUV) диапазоне. Эта информация важна на временных масштабах от дней до лет - для использования в различных моделях космической погоды, начиная от сопротивления спутника / затухания орбиты и заканчивая свойствами ионосферы;  3) потоки космических лучей на высотах коммерческой авиации, пилотируемой космонавтики (в том числе МКС) и спутников. Также энергичные частицы также блокируют радиосвязь в высоких широтах во время солнечных бурь.

Исследования космической погоды относится к прикладной области науки. Понимание и моделирование  явлений космической погоды затруднено разнообразными и сложными задействованными физическими механизмами, которые естественным образом не приводят к простому однозначному отображению. Целью фундаментальных исследований является понимание причин и развития конкретного природного явления, будь то солнечная вспышка, вызывающая радиовсплеск, изменение магнитного поля, вызывающее усиленные ионосферные токовые системы, или ионосферный электроджет, производящий мерцание, изготовление конструкций и повышение надежности технологических систем.

Наиболее продвинутым в области исследований КП являются США. В этой стране разработана цельная система наблюдений, анализа,  моделирования, обеспечивающая оперативное прогнозирование и выработку рекомендаций.  Официальным национальным источником оповещений о космической погоде в США является центр прогнозирования космической погоды (SWPC), который был создан в 2007 г. SWPC работает круглосуточно и без выходных. Развитие системы прогнозирования КП в нашей стране может опираться на опыт, накопленный в других странах, с учетом имеющихся систем предупреждения и прогнозирования опасных природных явлений, имеющихся наблюдательных средств, существующих научных и  информационных центров.

Термин "космическая погода", как правило, используется для описания «влияний» космической среды, а не физики самой космической среды. В настоящее время КП является болей общего направления исследований «гелиофизика»,  которая активно развивается и позволила понять многие явления солнечной и космической физики, которые определяют космическую среду.  Гелиофизика обычно делится на следующие три подобласти, имеющие отношение к космической погоде: физика Солнца и гелиосферы, которая охватывает Солнце и межпланетное пространство; физика магнитосферы, описывающая область межпланетного пространства, близкую к Земле, где физика все еще контролируется магнитным полем Земли; и физика верхних слоев атмосферы, которая сосредоточена на слоях атмосферы выше стратосферных уровней, включая мезосферу, термосферу и экзосферу, а также их ионизированные части, составляющие ионосферу.

В данной статье мы затронем только область гелиофизики и аспекты КП, связанные непосредственно с физикой Солнца и особенно наблюдениями солнечной активности.

## 2. Принципы построения прогнозирования космической погоды
### 2.1. Основные факторы КП

В настоящее время  основными компонентами в системе  прогнозирования космической погоды являются оценка высокоскоростных потоков солнечного ветра и прогноз геоэффективности корональных выбросов массы.

Из солнечной короны непрерывно истекает солнечный ветер (СВ) со средней скоростью ≈400 км/с. Солнечный ветер представляет собой плазму с вмороженным магнитным полем, потоки которой заполняют межпланетное пространство. Во время спокойных периодов, когда на Солнце нет возмущений вспышечного характера, приводящих к спорадическим выбросам в межпланетную среду ускоренной плазмы из короны, известных как корональные выбросы массы (КВМ), так называемый спокойный фоновый солнечный ветер формирует в межпланетном пространстве квазистационарную картину, определяющуюся, в основном, двумя факторами. Это – распределение, в зависимости от солнечных координат, начальной скорости истечения потоков из спокойной короны и вращение Солнца. В результате последующего радиального распространения потоков получается известная картина СВ с магнитным полем в виде архимедовой спирали, шаг которой определяется скоростью ветра в данном месте. Даже в спокойные периоды солнечной активности, скорость солнечного ветра может достигать ~700 км/с и выше. Это происходит, если потоки СВ формируются в корональных дырах и направлены к Земле. Такие высокоскоростные потоки солнечного ветра, от источников с открытой конфигурацией магнитных силовых линий, являются одним из основных факторов космической погоды. Оценка этих потоков, а также прогноз их интенсивности на несколько дней является предметом службы КП. Оценка этих потоков, а также определение "фоновых" параметров межпланетной плазмы является отправной точкой для моделирования распространения корональных выбросов массы. Характерные время изменения конфигурации крупномасштабных полей занимает несколько часов. Поэтому магнитографические наблюдения должны выполняться не реже чем несколько раз в сутки. Корональные выбросы массы являются одним из главных факторов формирования сильных геомагнитных возмущений. КВМ развиваются как в активных областях Солнца, например во время солнечных вспышек, так вне активных областей, например при эрупции спокойных волокон. Процесс выброса вещества занимает интервал от нескольких десятков минут до нескольких часов. Поэтому для детектирования КВМ и определения их параметров необходимы непрерывные круглосуточные наблюдения, со скважностью не хуже 5-15 минут.

Служба КП должна опираться на непрерывные наблюдения солнечной активности, а также включать в себя оперативный процесс первичного анализа данных и формирования индексов активности.

Данные наблюдений магнитных полей и солнечной активности задаются в качестве граничных условий в моделях распространения солнечного ветра и корональных выбросов массы (КВМ) через межпланетную среду. Параллельно с этим в прогноз КП могут включаться методики прогноз геоэффективности солнечных вспышек и КВМ.

Таким образом, в системе прогнозирования КП первоочередными источниками являются информация о рекуррентных потоках солнечного ветра и данные о корональных выбросах массы. Для получения информации об этих процессах используются различные наблюдательные средства. Для оценки параметров солнечного ветра используются магнитографы полного диска Солнца. Для регистрации КВМ необходимы непрерывные наблюдения хромосферы или ближней короны Солнца.

## 2.2. Наблюдения магнитных полей и моделирование рекуррентных потоков СВ

Солнечный ветер формируется под действием корональных магнитных полей на Солнце. Наиболее значимыми для формирования СВ являются относительно медленно меняющиеся крупномасштабные магнитные поля на солнечной поверхности. В настоящее время наблюдения таких полей осуществляется с помощью солнечных магнитографов фотосферного поля на полном диске Солнца. Наиболее подходящим из отечественных приборов для наблюдений крупномасштабных магнитных полей является СТОП (Солнечный телескоп оперативных прогнозов) поколения (Пещеров и др. 2013). Он разработан, на основе глубокой модернизации работающего в Саянской солнечной обсерватории прежнего варианта СТОП, в ИСЗФ СО РАН по заказу Росгидромета в рамках ФЦП «Геофизика». Изготовлено три экземпляра таких телескопов, которые установлены в обсерваториях: Байкальской обсерватории (ИСЗФ, Иркутск), Уссурийской обсерватории и на Кисловодской Горной станции ГАО (ГАС ГАО). В 2014 г. на наблюдательной площадке ГАС ГАО начались регулярные наблюдения на магнитографе СТОП (Tlatov et al. 2015a). Эти наблюдения без перерывов продолжаются и до настоящего времени. В Уссурийске и Иркутске регулярные мониторинговые наблюдения на магнитографе наладить не удалось.

Фотосферное магнитное поле Солнца обычно наблюдают с помощью эффекта Зеемана. Взаимодействие между магнитным моментом электрона с внешним магнитным полем расщепляет атомные энергетические уровни. Спектральная линия расщепляется на две компоненты, и каждая компонента будет иметь свое состояние поляризации. В присутствии магнитного поля вдоль луча наблюдения (LOS) наблюдателя, спектральная линия расщепляется на две сдвинутые компоненты σ имеющих круговую поляризацию и несмещенную компоненту π. Сдвиг длины волны Δλ пропорционален полю B$l$, Δλ~ B$l$. Это так называемый продольный эффект Зеемана. В присутствии магнитного поля перпендикулярно к лучу зрения (B$t$), смещенные компоненты линейно поляризованы. Этот явление носит называние поперечного эффекта Зеемана. На практике профили поляризованных спектральных линий описывают параметрами Стокса(I,Q,U,V). Параметр Стокса I($\lambda$), характеризует интенсивность в зависимости от длины волны. Параметр Q($\lambda$) разница интенсивности между вертикальной и горизонтальной компонентами линейной поляризации. Параметр U($\lambda$) разница интенсивности линейной поляризации на углах +45 и -45$^o$. Параметр V($\lambda$), разница интенсивностей между правой и левой круговой поляризацией. Линейно поляризованный сигнал (Q,U), как правило, на порядок слабее круговой поляризации (V). Как правило, предел чувствительности измерения компоненты B$t$ порядка 100 G, в то время как чувствительность инструментов для компоненты B$l$ ~1-10G.
Для моделирования крупномасштабного магнитного поля в короне необходимо измерение слабых крупномасштабных полей, величиной порядка несколько единиц Гаусс, поэтому для этих целей могут подходить наблюдения только компоненты B$l$. На Рис.1 представлены распределения крупномасштабных полей Солнца по данным наблюдений магнитографа СТОП и наблюдательной сети GONG. Для лучшего сравнения слабых полей приведена величина B$^{0.5}$. Для устранения годовой вариации данные GONG экстраполированы в высоких широтах. Мы видим хорошее соответствие между двумя видами наблюдательных данных. Вместе с тем, видны и различия. Так в период минимума активности 2019-2020 гг. Вблизи экватора видно различие, связанные со сдвигом нуля в данных GONG. Даже такие малые различия могут существенно повлиять на результаты моделирования коронального магнитного поля. Следует отметить, что магнитографы СТОП и GONG используют различные схемы. СТОП это магнитограф на базе спектрографа, а GONG фильтровый магнитограф.

Для создания полной синоптической карты магнитного поля мы выбираем наилучшую карту для каждого дня наблюдений, по критерию наименьшей средней ошибки наблюдений на солнечном диске. Определялся список дней наблюдений, который мог давать вклад в данные для выбранной синоптической карты. Затем для каждой долготы синоптической карты и таблицы дней наблюдений определялась весовая функция в виде w=1/Δt2, где Δt – разница между долготой на синоптической карте и долготой центрального меридиана на текущем дне наблюдений. Дни наблюдений, отстоящие от центрального меридиана на долготу более 60 градусов, не учитывались. Сумма весовых функций для всех дней наблюдений k, дающих вклад на данной долготе от номера l до m должна быть равна 1: Реконструкция осуществлялась с учетом угла наклона солнечного экватора b. Созданная синоптическая карта из данных наблюдений вдоль луча зрения депроектировалась в предположении радиального магнитного поля на фотосфере (Harvey and Worden, 1998;Wang and Sheeley, 1992).

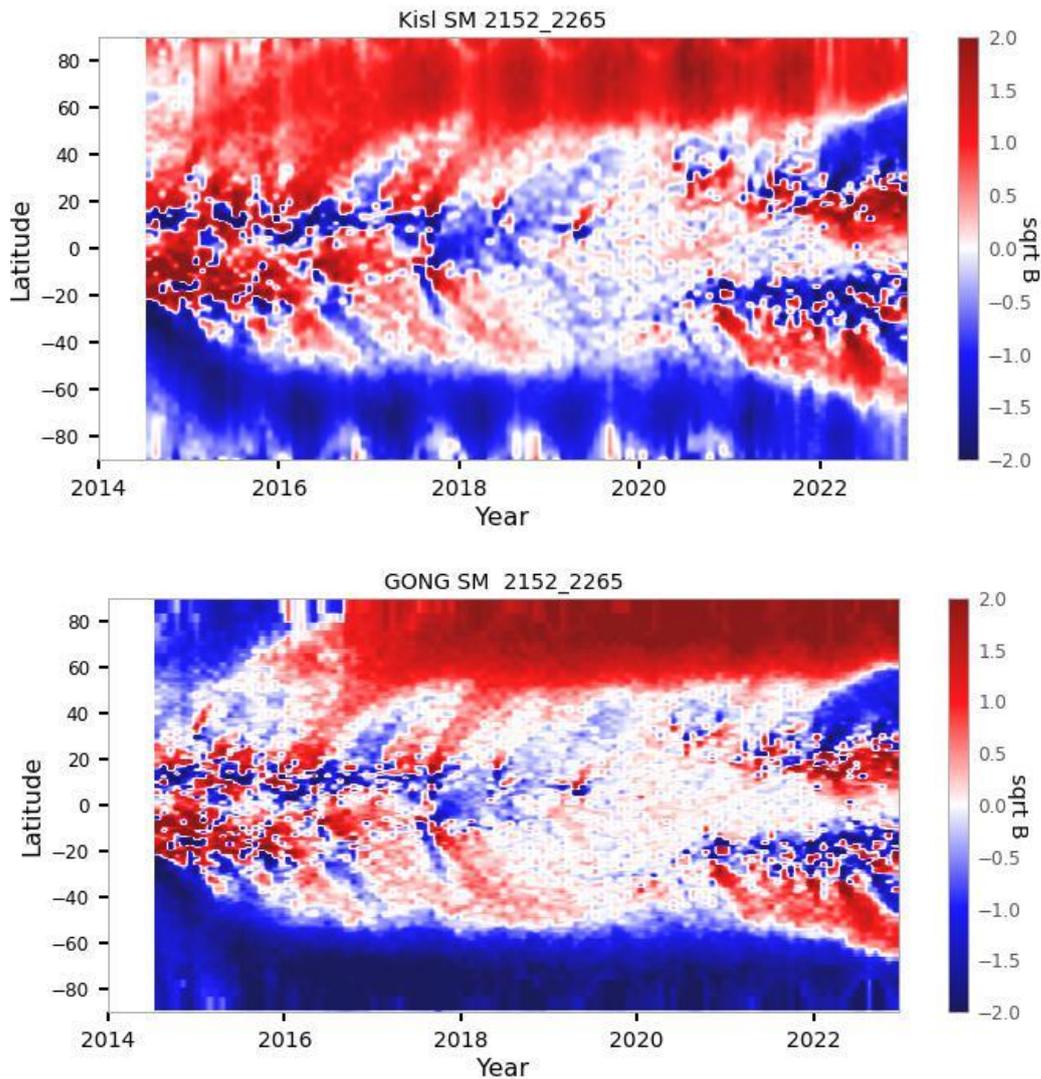

**Рис. 1**. Широтно-временная диаграмма распределения магнитных полей на Солнце. a) по данным наблюдения магнитографа СТОП в Кисловодске; b) по данным наблюдений сете GONG.

Для моделирования глобальной магнитного поля в короне широко используется модель потенциального магнитного поля с поверхностью источника (PFSS). В этой модели используется фотосферное магнитное поле в виде синоптической карты как граничное условие на внутренней границе R=1. Модель использует предположение, что на гипотетической поверхности источника на расстоянии 2.5R от центра Солнца, силовые линии магнитного поля считаются радиальными (Schatten, Wilcox, and Ness, 1969).

Хотя механизм ускорения СВ еще не полностью понят, на протяжении последних десятилетий была подтверждена корреляция между геометрией коронального поля и скорости SW. Солнечный ветер происходит из открытых структур магнитного поля в короне и ускоряет вдоль открытых силовых трубок (Parker, 1961). Их основным источником являются области, известные как корональные дыры (КД), которые имеют более низкую плотность и температуру, чем фоновая корона. Магнитное поле в КД преимущественно униполярное. Форма, размеры и расположение КД значительно меняются с течением цикла активности Солнца. Структура изменения солнечного ветра соответственно тоже. Wang and Sheeley (1990) используя синоптические карты магнитного поля и модель PFSS, вычислили коэффициенты расширения магнитных силовых трубок (FTE). Они обнаружили существенную обратную корреляцию между параметром FTE и скоростью солнечного ветра, наблюдаемого в плоскости эклиптики. Фактор расширения магнитного поля fs определяется в каждой точке P поверхности источника при сравнении вычисленной интенсивности $B^p(Rs)$ с ассоциированным значением на фотосфере $B^p(Ro)$, получаемый при движении вдоль силовой линии. Математически:

$$f_s = (R_\Theta / R_S)^2 \left[ B^P(R_\Theta) / B^P(R_s) \right], \quad (1)$$

где Br(Ro) и Br(Rs) напряженность магнитного поля на поверхности Солнца и на поверхности источника в одной и той же силовой трубке. Arge and Pizzo (2000) предложили эмпирическую функцию для связи между FTE и скоростью солнечного ветра. Такая функция давала существенную корреляцию (~0,4) между моделью и наблюдаемой скорости солнечного ветра за период около 3 лет. Arge et al. (2004) позднее изменили эмпирическую функцию, добавив дополнительный параметр (θb), который описывает минимальное угловое расстояние на фотосфере между основаниями открытых силовых линий границей ближайшей границей корональной дыры. Силовые трубки из центра больших полярных КД имеют большие углы θb и, следовательно, производят более высокую скорость. Как показывает анализ (Berezin, Tlatov, 2022), параметр θb имеет гораздо большую прогностическую ценность, чем фактор расширения fs.

После кинематического распространения СВ от Солнца до орбиты Земли, можно сделать прогноз скорости и полярности IMF (по отношению к направлению на Солнце) на орбите Земли за несколько дней. Эта модель, известная как модель Ван-Шили-Арге (WSA), в настоящее время широко используется для прогнозирования космической погоды. Исходными данными для этой модели служат синоптические карты магнитного поля Солнца, полученные на основе магнитограмм, для обеспечения, зависящего от времени описания фоновой плазмы солнечного ветра и межпланетного магнитного поля (Рис. 2). Эмпирическая функция, предложенная Arge et al. (2004) выражается в форме:

$$V(f_s, \theta_b) = 265 + \frac{1.5}{(1+f_s)^p} \left\{ a - b e^{-(\theta_b/\theta_0)^q} \right\}^k \text{ km s}^{-1}, \quad (2)$$

Для данных магнитографа СТОП соотношение было оптимизировано (Berezin, Tlatov, 2022):

$$V(f_s, \theta_b) = 265 + \frac{3.4}{(1+f_s)^{0.03}} \left( 3.14 - 2.97 e^{-(\theta_b/2.5)^{0.47}} \right)^{6.5} \text{ km s}^{-1} \quad (3)$$

Выходные данные о СВ затем передаются в связанную магнитогидродинамическую (МГД) модель, например, ENLIL. Или в другие модели распространения СВ, например, в баллистическую. модель распространения потоков солнечного ветра от поверхности источников, как это происходит на ГАС (http://solarstation.ru/sun-service/forecast). При этом быстрые и медленные потоки СВ могут взаимодействовать между собой. На Рис. 3 представлен пример оперативного прогноза параметров СВ на орбите Земли, выполненного по данным телескопа СТОП. Данные прогноза оперативно обновляются на сайте (http://solarstation.ru/sun-service/forecast).

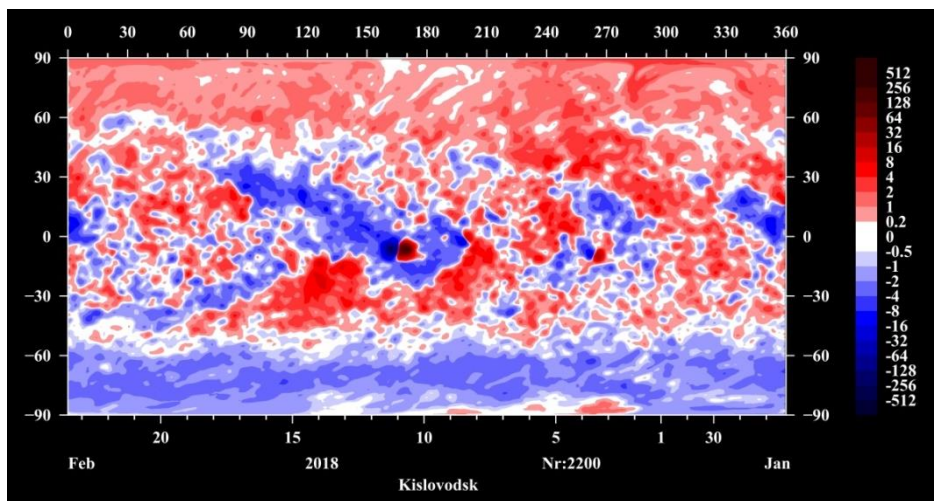

**Рис. 2**. Синоптическая карта распределения магнитных полей по данным наблюдений магнитографа СТОП.

Модельные расчёты скорости СВ по данным магнитографов СТОП, WSO, GONG, SDO/HMI демонстрируют достаточно точное соответствие с данными измерений на спутнике ACE в определённые периоды (Рис. 4) (Berezin, Tlatov, 2021). Расхождения между моделью и наблюдениями могут быть связаны, с возмущениями солнечного ветра и с ошибками наблюдения магнитного поля, особенно на полюсах (Jian et al., 2011). Среди рассмотренных наземных приборов WSO и СТОП обеспечивают наиболее высокую точность расчета, как параметров солнечного ветра, так и параметров корональных дыр. Кроме того, за рассмотренный период (2014 – 2019 гг.) наземные измерения WSO и СТОП показывают, в среднем, лучшие результаты в расчёте скорости спокойного солнечного ветра относительно спутниковых измерений (SDO/HMI) при одинаковых параметрах модели. Наименее точную информацию, с отрицательной корреляцией за 2015 г., предоставляют данные сети GONG, однако этот результат может быть несколько улучшен при помощи надлежащей полярной коррекции.

Систематические отклонения в амплитуде модельного спокойного солнечного ветра могут быть связаны с временной изменчивостью эмпирических коэффициентов модели WSA, в частности, коэффициента q (формула 1). Выявление причин указанной изменчивости позволит добиться повышения качества прогноза.

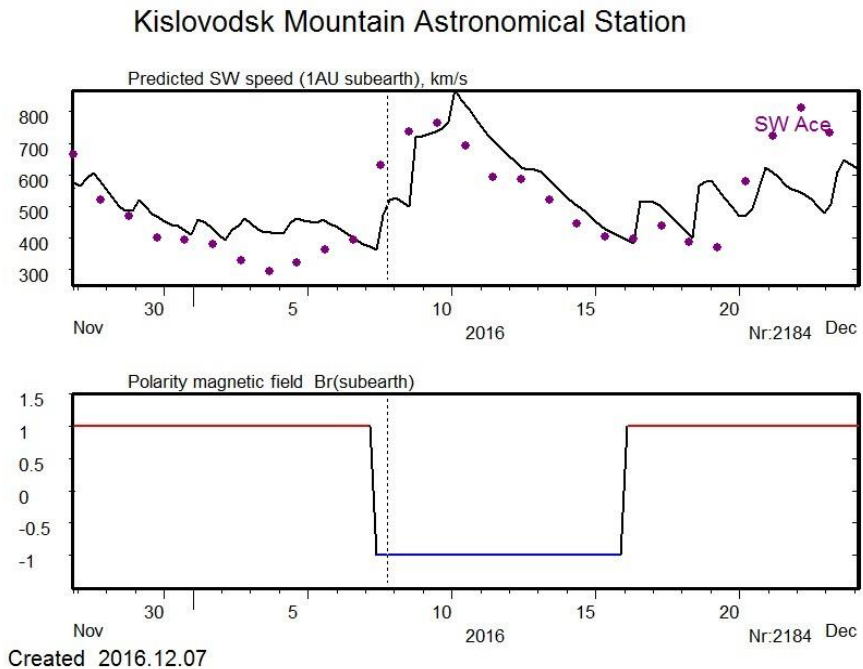

**Рис. 3**. Пример оперативного прогноза скорости и полярности солнечного ветра на ГАС, выполненного 07.12.2016 и его сравнение с данными наблюдений обсерватории ACE. Пунктирной линией отмечена дата прогноза.

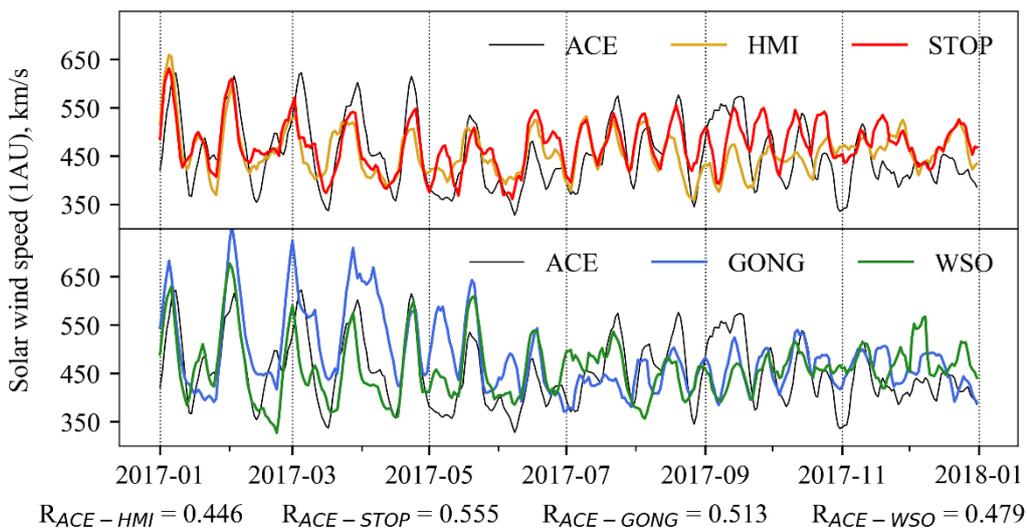

**Рис. 4**. Спутниковые измерения скорости солнечного ветра (ACE) и модельные расчёты в 2017 году по данным различных магнитографов (Berezin, Tlatov, 2021).

Таким образом, эта составляющая прогноза КП в нашей стране принципиально решена при использовании уже имеющихся наблюдательных средств, хотя и требует дальнейшего совершенствования.

### 2.3. Детектирование, определение параметров и прогноз прибытия КВМ

Корональные выбросы массы являются основным фактором, определяющим сильные геомагнитные бури. Корональные выбросы массы— это обширные выбросы намагниченной плазмы из короны Солнца. На Земле КВМ являются основным фактором сильных геомагнитных возмущений и космической погоды, которая питает космическую среду Земли и нарушает работу критически важных услуг, предоставляемых космическими кораблями, энергосистемами и самолетами (Cannon et al., 2013; Hapgood, 2011). КВМ распространяются в солнечном ветре. Понимание распространения КВМ в солнечном ветре и возможность оценить их ожидаемое прибытие на Землю являются ключевыми исследовательскими вопросами и задачами для центров прогнозирования космической погоды. Несмотря на недавний прогресс, эволюция КВМ через солнечный ветер и гелиосферу до сих пор недостаточно хорошо изучена из-за редких гелиосферных наблюдений и открытых вопросов относительно структуры КВМ (Luhmann et al., 2020).

Для успешного прогнозирования КП необходимы регулярные наблюдения КВМ. В США NOAA/SWPC такие наблюдения проводились и проводятся на космических коронографах. Одним из наиболее востребованных является и широкоугольной спектрометрический коронограф белого света (LASCO) в рамках миссии NASA/SOHO (Brueckner et al., 1995) До 2007 г. корональные выбросы можно было регулярно наблюдать только вблизи Солнца (обычно в пределах 30 солнечных радиусов). С 2007 с момента запуска миссии STEREO (Kaiser et al., 2008) появилась возможность наблюдать за распространением КВМ из солнечной короны через внутреннюю гелиосферу на околоземную орбиту, используя коронограф белого света и гелиосферный формирователь изображений (HI) в пакете инструментов Sun-Earth-Connection-Heliospheric-Investigation (Howard et al., 2008).

В нашей стране космических солнечных обсерваторий не существует, и их создание займет, очевидно, долгий период. Таким образом, наблюдения и детектирование КВМ является критически важной технологией, которую нужно освоить в кратчайшее время. Для этой цели могут использоваться наземные оптические (Tlatov et al. 2017), и радиотелескопы (Grechnev et. al. 2006), регистрирующие процессы на полном диске Солнца и работающие в непрерывном режиме. В данной работе мы рассмотрим опыт работы оптических телескопов. Наблюдение, детектирование и определение параметров КВМ может осуществляться на наземных патрульных телескопах. Такие телескопы обеспечивают непрерывные наблюдения полного диска Солнца с периодичностью порядка минуты в хромосферные линиях Н-альфа и CaIIK. Прототипы таких телескопов были созданы на ГАС ГАО. Автоматизированной телескоп-спектрогелиограф SPOT (Солнечный Патрульный Оптический Телескоп), работающий в линии CaIIK установлен и осуществляет регулярные наблюдения, на ГАС ГАО с 2012 г. (Tlatov et al., 2015b). В 2015 г. создан патрульный автоматический спектрогелиограф, осуществляющий непрерывные наблюдения в линии Н-альфа (Рис. 5). В настоящее время, на ГАС получен опыт длительных наблюдений на спектрогелиографах.

Созданные на ГАС патрульные телескопы имеют различную конструкцию. Патрульный телескоп в линии CaIIK выполнен в одно объёмной схеме. Ось телескопа направлена на мировую, и течение наблюдений весь телескоп, включая спектрограф вращается. Сканирование по щели спектрографа осуществляется магнитными подвижными элементами. Телескоп в линии Н-альфа выполнен по стандартной схеме спектрогелиографа. Сопровождение Солнца осуществляется Йенш-целостатом. В этом целостате основное и перекидное зеркала установлены на одной подвижной колоне. Это схема позволяет наблюдать в течение длительного времени, при котором зеркала не заслоняют друг друга. Скорость вращения зеркал различается примерно в 2 раза. Далее пучок света подается на телескоп. Телескоп выполнен как рефрактор с диаметром объектива 10 см и фокусным расстоянием 150 см. Для сканирования Солнца по щели используется подвижная платформа с 2-мя плоскими зеркалами и возможностью перемещения по двум направлениям, для юстировки фокуса и сканирования. Создание нового функционального элемента для наблюдения Солнца в линии Н-альфа имело ряд оригинальных конструкторских решений, которые были оформлены как Патент на полезную модель N 126854 от 10 апреля 2013 г. (Рис. 2.5).

Время сканирования диска Солнца составляет от 50 до 120 секунд в зависимости от времени экспозиции и времени, необходимого для позиционирования Солнца на фотогиде. Количество, кадров при перемещении Солнца через спектральную щель составляет ~2000.

Для регистрации спектра используется CCD-матрица Prosilica GT 3300 с разрядностью ADC 14 bit. Спектральное разрешение ≈ 40000, линейная дисперсия ≈ 0.16 Å / пиксель, интервал дискретизации спектра ≈ 0.03 Å / пиксель. Максимальное количество наблюдений полного солнечного диска может достигать 450 изображений в день, что соответствует 6.5-8 часам непрерывных наблюдений.

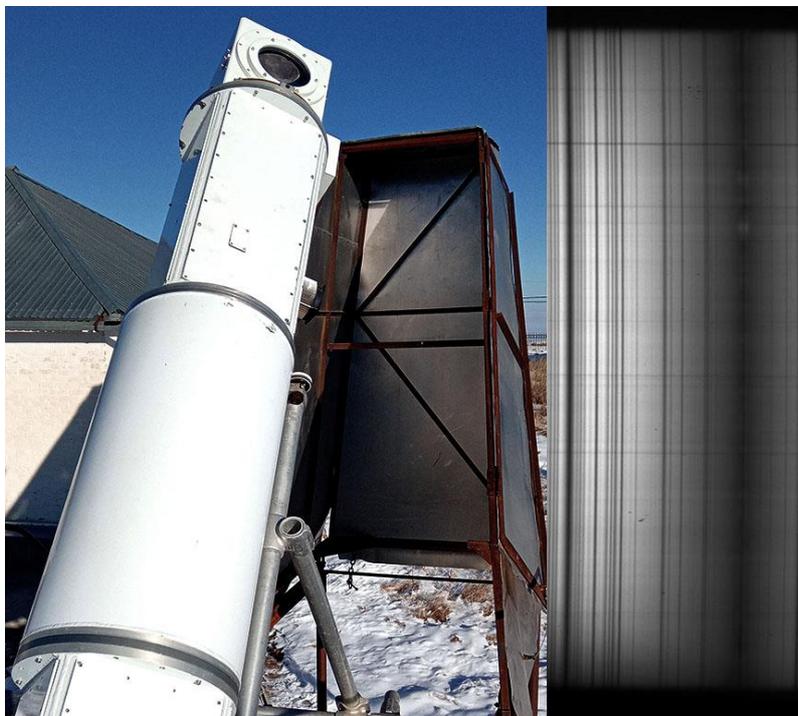

**Рис. 5**. Слева: патрульный полярно-осевой телескоп линии Ca II K, справа: пример спектров, наблюдаемых в одном положении щели спектрографа.

Результатом наблюдений патрульных телескопов является серия кадров спектров солнечного света, получаемых при прохождении Солнца через щель спектрографа. На Рис. 6 представлен пример кадра спектра Солнца вблизи линии H-альфа. Линия H-альфа видна здесь как темная область. Ширина спектра, его положение и интенсивность меняются, в зависимости от объектов расположенных на Солнце. Также на спектре видны темные области, распространяющиеся в континуум. Как правило, это солнечные пятна. Размеры кадра по горизонтали $x$ и вертикали $y$ составляют 2000x70 пикселей. На кадре можно отметить кривизну спектральных линий. Это связано с тем, что при работе в высоких порядках спектра пучки света от щели идут под углом к главной плоскости, что приводит к кривизне спектральных линий. Кривизна спектральных линий может быть компенсирована с помощью изогнутой входной щели (например, VSM/SOLIS), однако в данном случае использована прямая щель спектрографа и последующая цифровая коррекция кривизны как более простое и дешёвое решение.

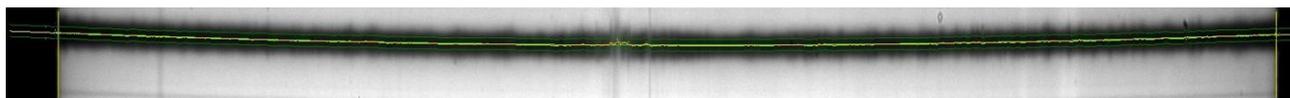

**Рис. 6**. Пример кадра спектра Солнца вблизи линии H-альфа. Проведены границы диска Солнца, точки центров аппроксимацией гауссианой распределения (зеленные точки), линия аппроксимации положений центров гауссиан полиномом второго порядка (красная линия).

Опишем основные этапы обработки спектральных данных на примере данных патрульного телескопа в линии H-альфа. При обработке данных наблюдений первым этапом является определение искривления спектральных линий. Для этого на каждом спектральном кадре определяются границы солнца по градиенту интенсивности, затем в каждом столбце кадра находятся центры линии в первом приближении, как точки наименьшей интенсивности. Полученные центры линии аппроксимируются полиномом второй степени, далее относительно полученной линии аппроксимации отбираются профили линии H-альфа фиксированной ширины. На втором этапе обработки определяются параметры всех отобранных профилей (ширина, амплитуда, положение центра) посредством аппроксимации профилей функцией Гаусса. На последнем этапе выполняется аппроксимация положения центров линии полиномом второй степени. Относительно неё вычисляются доплеровские смещения.

В результате обработки кадров спектра получаются изображения различные изображения в том числе: изображения в центре линии, распределение лучевых скоростей и ширины линии в каждой точке на диске Солнца и хромосфере (Рис. 7). Яркие области (Рис. 7, верхняя левая панель) соответствуют хромосферным флоккулам, связанным с АО 12680 (около лимба на востоке); АО 12678 и 12677 (к западу от центрального меридиана); АО 12679, 12674, 12673 (около лимба на западе). Тёмные области (к югу от экватора) соответствуют волокнам, как и на изображении в линии H-альфа сети GONG (Рис.6, нижняя правая панель). Флоккулы также видны в распределении ширины линии как области, в которых спектральный профиль H-альфа относительно широкий (Рис. 7, нижняя левая панель). Это происходит из-за более высоких температур. На внешних границах флоккул спектральные профили более узкие, что соответствует более низким температурам.

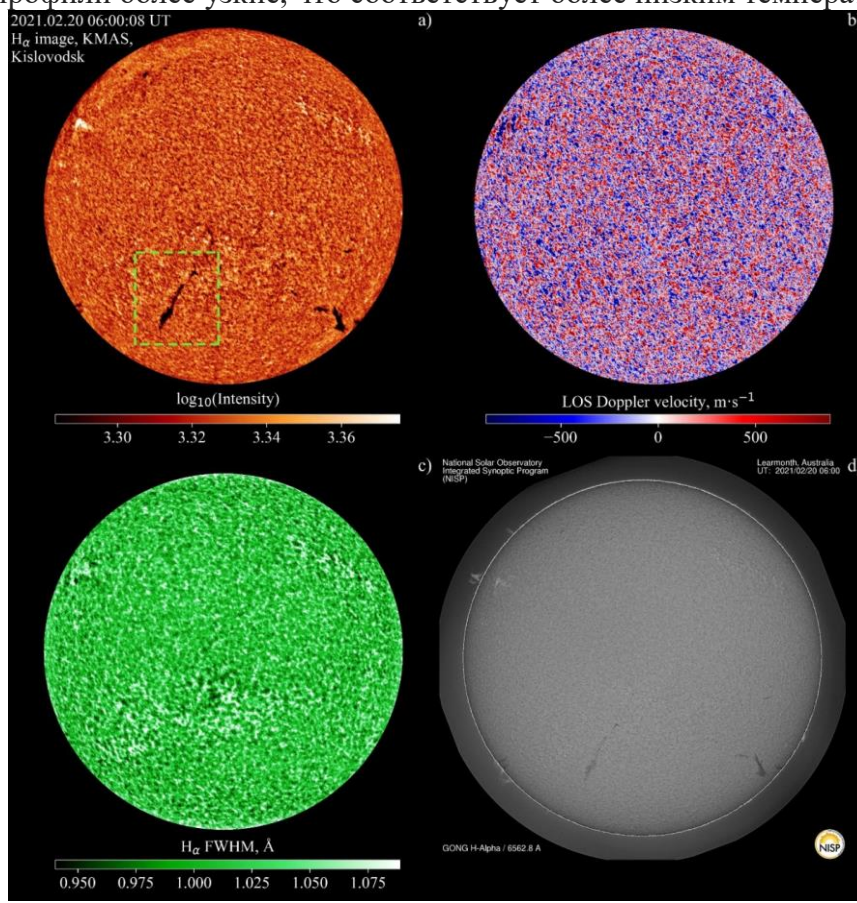

**Рис. 7**. Пример восстановленных карт интенсивности в центре линии H-альфа (верхняя левая панель), доплеровских скоростей (верхняя правая панель) и ширины H-альфа (нижняя левая панель) по данным наблюдений 20 Февраля 2021 в 06:00 UT UT. Зелёным контуром на карте доплеровских скоростей обозначены границы ярких элементов. На правой нижней панели приведено изображение в линии H-альфа Global Oscillations Network Group (GONG), полученное в 06:00 UT. (Berezin et al., 2023).

Выбор схемы спектрогелиографов, с регистрацией полного профиля спектральной линии, в качестве схемы таких телескопов, позволяет существенно повысить контраст регистрируемых эруптивных событий. На Рис. 8 представлен пример изображения Солнца в ядре линии CaIIK во время эруптивного события 02.10.2014. Также представлена разность интенсивностей в правом и левом крыле спектральной линии. На этой диаграмме в северо-восточной части диска видны в два темных эруптивных выброса, эволюцию которых удалось проследить на протяжении около 40 минут. Заметим, что контраст эруптивных событий, получаемых методом вычитания интенсивности в крыльях линии значительно выше, чем методом вычитания изображений в разное время методом EUV димминга. Патрульные телескопы позволяют также регистрировать эруптивные события на лимбе Солнца и отслеживать их до высот порядка 1.3-1.4Ro.

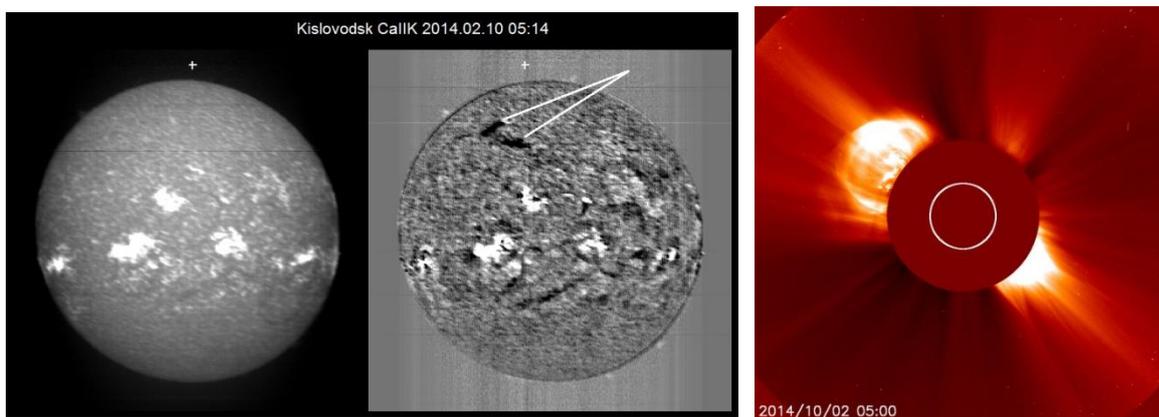

**Рис. 8**. Изображения Солнца в ядре линии CaIIK (слева) во время эруптивного события 02.10.2014. Также представлена разность интенсивностей в правом и левом крыле спектральной линии (посередине). Стрелками указаны два эруптивных выброса. (Справа) Развитие КВМ по данным LASO.

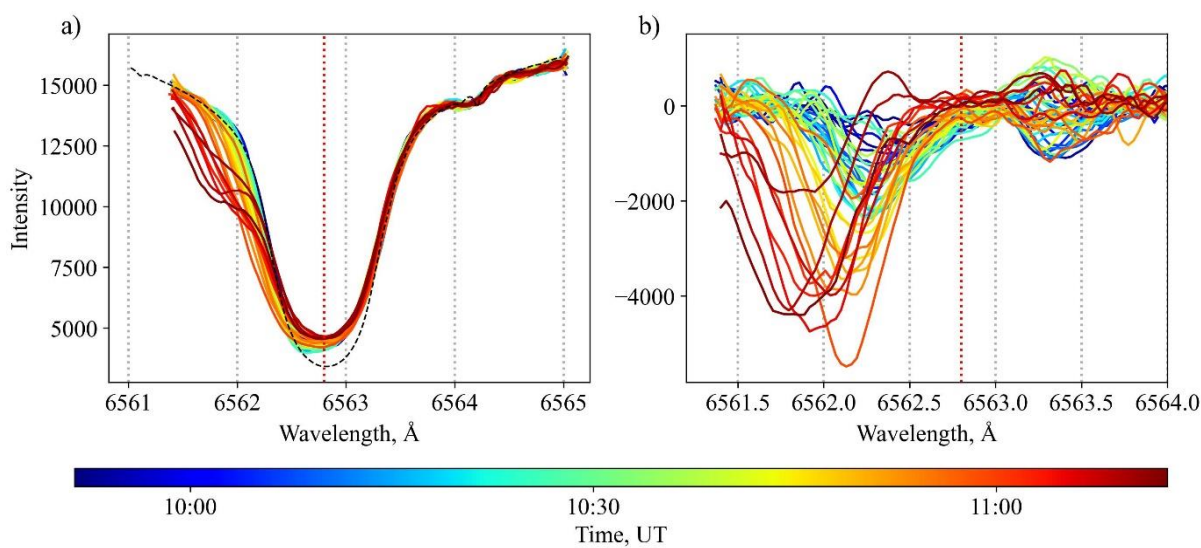

**Рис. 9**. Эволюция спектрального профиля H$_\alpha$ в центральной точке волокна в процессе эрупции 20 февраля 2021 г.. На панели (a) представлен наблюдаемый спектральный профиль, на панели (b) - разность между наблюдаемым профилем и профилем спокойной хромосферы (Berezin et al., 2023).

Рассмотрим процесс регистрации эрупции волокна на патрульном спектрогелиографе ГАС в линии Hα на примере события 20.02.2021 (Рис.7). Серия наблюдений 20 февраля в этот день началась за несколько часов до КВМ. На Рис. 9а представлена эволюция наблюдаемого спектрального профиля в центральной точке волокна на финальной стадии эрупции. Мы видим рост возмущений в синем крыле Hα, связанный с быстрыми восходящими потоками в волокне. После вычитания спектрального профиля, характерного для спокойной хромосферы, мы можем видеть возрастающие допплеровские синие смещения, которые достигают 1 Å, что соответствует скорости ~46 км/с. Согласно базе данных HEK, по наблюдениям в ультрафиолете (SDO/AIA304) эрупция началась в 9:30 UT. Наш анализ скорости во всех ассоциируемых с волокном точках показывает, что началу эрупции предшествует фаза медленного роста средней скорости до нескольких км/с (Рис.10d). Затем ускорение растёт, и на финальной стадии средняя скорость достигает ~45 км/с. В процессе наблюдения волокно постепенно исчезает из ядра Hα (Рис. 10a) и более контрастно проявляется в синих смещениях (Рис.10b). В фазе медленного подъёма распределение скоростей близко к симметричному, восходящие движения преобладают, но наблюдаются и нисходящие потоки (Рис.10с 09:10 UT, 09:40 UT). Затем распределение скоростей становится более ассиметричным, преобладают вариации скорости в относительно медленных потоках – до ~25 км/с (Рис.10с 10:10 UT).

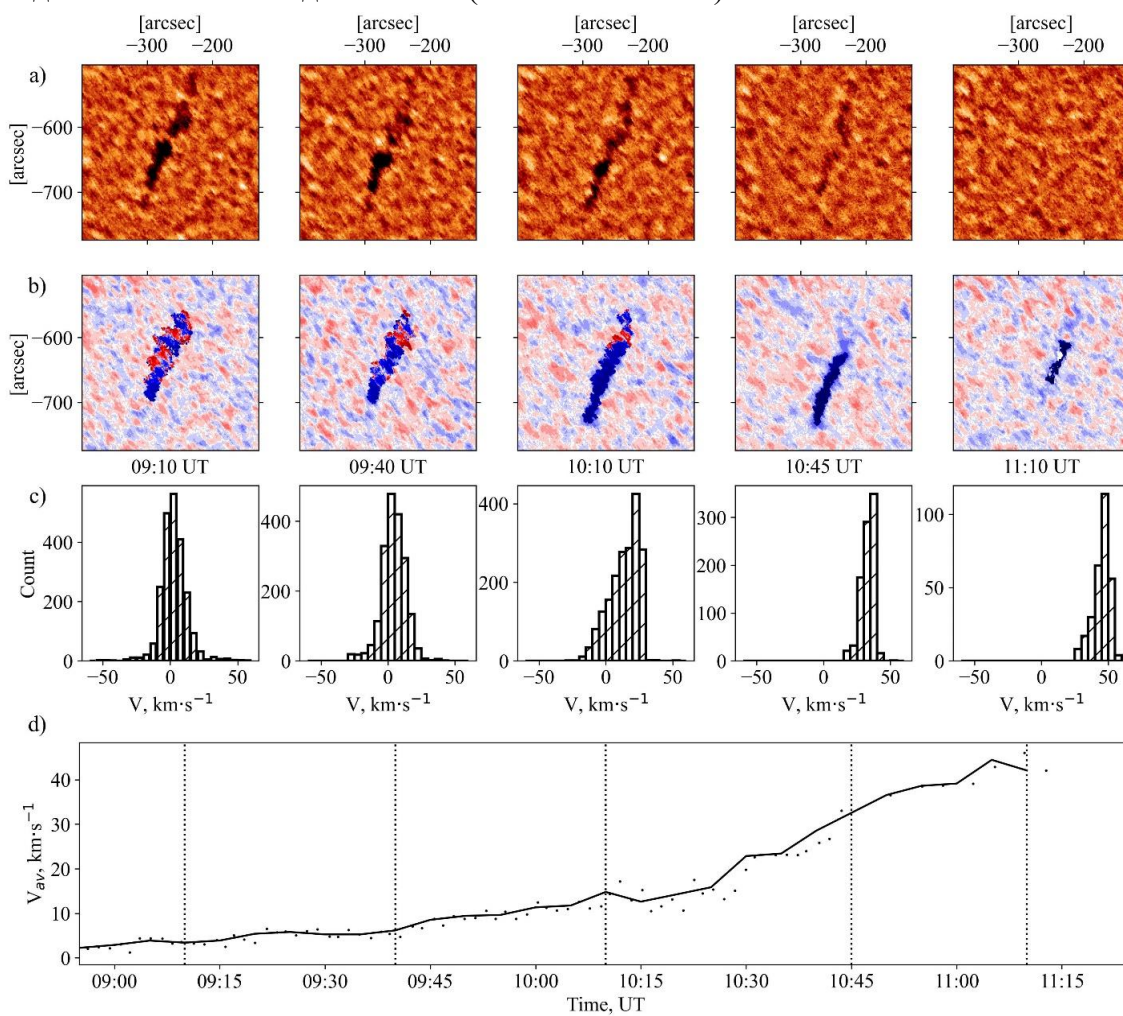

**Рис. 10**. Наблюдение процесса эрупции волокна 20 февраля 2021 г. Панель (a): изображения в ядре Hα в разные моменты времени. Панель (b): LOS скорости в волокне (V) и в спокойной хромосфере; синий и красный цвет соответствует синим и красным смещениям соответственно. Гистограммы распределения скорости V в волокне представлены на панели (c); отрицательные и положительные значения скорости соответствуют нисходящим и восходящим потокам соответственно. Панель (d): средняя по волокну LOS скорость Vav (Berezin et al., 2023).

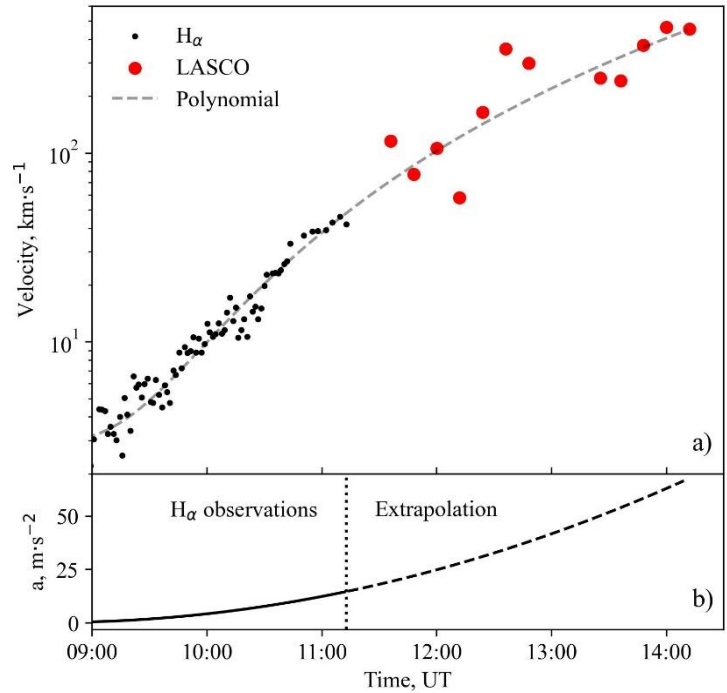

**Рис. 11**. Средняя скорость восходящего движения эруптивного волокна в направлении LOS; сопоставление со скоростью наблюдаемого на коронографе LASCO КВМ 20 февраля 2021 г. (a). Оценка ускорения КВМ a по наблюдениям в линии $H_\alpha$ начального этапа КВМ (b). (Berezin et al., 2023).

На финальной стадии эрупции мы наблюдаем только восходящие потоки, которые обладают скоростями из относительно узкого диапазона значений (Рис.10c 10:45 UT, 11:10 UT).

Около 11:30 UT КВМ был зафиксирован на коронографе SOHO/LASCO. Мы сопоставили наши оценки средней начальной скорости эруптивного волокна с наблюдениями на коронографе (Рис. 11a). На Рис. 11a также представлена аппроксимация скорости, полученной из наблюдений $H_\alpha$, полиномом третьего порядка. Отчётливо видно, что более поздние наблюдения на коронографе хорошо ложатся на эту аппроксимирующую кривую. Это позволяет нам сделать вывод о том, что патрульные наблюдения на спектрогелиографе в линии $H_\alpha$ можно использовать для оценки начального ускорения КВМ. В процессе наблюдения начальной стадии эрупции ускорение КВМ растёт до ~13 м/с$^2$, дальнейшая экстраполяция показывает рост ускорения до значений более 60 м/с$^2$, что подтверждается наблюдениями на коронографе (Рис. 11b).

Средняя скорость волокна меняется по закону, близкому к экспоненциальному, хотя мы аппроксимируем скорость полиномом. Дальнейшая экстраполяция скорости, в целом, совпадает с наблюдениями на коронографе LASCO (Рис. 10a).

Важным параметром сети наблюдений за КВМ является возможность непрерывного режима наблюдений. Наземная сеть GONG, состоящая их 6 наблюдательных пунктов, разнесенных по долготе, включающая телескопы в линии H-альфа с периодичностью съемки 1 минута обеспечивает покрытие не менее 90% процентов времени суток [Jain et al., 2021]. Это соизмеримо с космической платформой на геостационарной орбите SDO. Мы можем ожидать, что сеть патрульных телескопов в количестве 6- 8 единиц обеспечит достаточное перекрытие для регистрации большинства КВМ. Таким образом, мы можем заключить, что патрульные наблюдения на спектрогелиографе в линии $H_\alpha$ и CaIIK могут использоваться для детектирования эрупции и КВМ и оценки начальной скорости и, возможно, массы КВМ.

## 3. Структура наблюдательной сети обновленной службы Солнца.

Национальная наблюдательная сеть Службы Солнца может быть создана на основе наземных телескопов для непрерывных синоптических наблюдений солнечной активности. Для успешного функционирования службы прогноза КП, также нужно обеспечить разработку методик анализа и прогнозирования состояния космической погоды (КП) на основе получаемых данных. Такая сеть может поставлять следующие типы прогнозов:

- прогноз параметров фонового солнечного ветра на период до 7 дней на расстоянии до 1.3 а.е. от Солнца;
- детектирование и прогноз геоэффективности корональных выбросов массы (КВМ) с заблаговременностью 2-3 дня до прихода к Земле;
- регистрация места, времени и мощности солнечных вспышек; оценка потоков энергичных частиц.
- оценка уровня ультрафиолетового излучения во время солнечных вспышек и спокойные периоды;
- прогноз вероятности и относительной амплитуды солнечных вспышек от 1 часа до 2 дней;
- прогноз продолжительных периодов высокой и низкой активности;
- прогноз солнечной эрупции и КВМ, с заблаговременностью до 10 часов.

Наземная наблюдательная сеть должна обеспечивать:
a) Наблюдения крупномасштабных магнитных полей Солнца определения параметров солнечного ветра и крупномасштабных возмущений межпланетного магнитного поля (ММП) не реже 1 раз в сутки.
b) Регистрацию эруптивных процессов и корональных выбросов массы на начальном стадии движения оптическими спектрогелиографами и (возможно) радиогелиографом ССРТ.
c) Патруль хромосферы – вспышки, волокна, оценку потоков УФ и жесткого излучения. Режим наблюдений непрерывный 24 часа, с интервалом не хуже раз в 1-2 минуты.
d) Картирование Солнца (и ближней короны) в оптическом и сантиметровом диапазонах на ряде фиксированных длин волн с измерением яркостных и поляризационных характеристик, расщеплений и сдвигов линий, что позволяет, в конечном счете, получать информацию о структуре движений, магнитных полях, оценку потоков жесткого УФ излучения.
e) Патруль радиовсплесков с определением их спектрального типа в непрерывном режиме.
Служба наблюдения за солнечной активностью может состоять из нескольких компонентов и пунктов наблюдений:
1. Наземная наблюдательная сеть может состоять из трех основных наблюдательных пунктов разнесенных по долготе в Уссурийске, Иркутске, Кисловодске. Эти наблюдательные пункты должны быть оснащены : магнитографом СТОП".
2. Сеть оптических патрульных телескопов, выполняющий наблюдения в линиях H-альфа и CaIIK. Телескопы предназначены для детектирования и определения скорости КВМ на начальном этапе разгона. Также данные используются для регистрации солнечных вспышек и оценки потоков УФ излучения. Такие автоматические телескопы могут быть установлены Уссурийске, Иркутске, Кисловодске и КрАО. Также целесообразно установка автоматических патрульных телескопов в других наблюдательных пунктах, в том числе в Западном полушарии.
3. Радиотелескопы для непрерывной регистрации и прогноза солнечных вспышек на волнах 10,7 и 5 см., ведущий наблюдения в автоматическом режиме. Такие телескопы уже существуют в Уссурийске, Иркутске, Кисловодске, КрАО. Целесообразна установка радиотелескопов в пунктах в западном полушарии.
4. Саянский солнечный радиогелиограф (ССРТ) для регистрации эруптивных событий и солнечных вспышек. В настоящее время ССРТ находится в стадии строительства и его эффективность для целей прогнозирования КП не подтверждена.
5. Телескоп регистрации полного вектора магнитного поля СОЛСИТ (Иркутск). Телескоп позволит проводить исследованию по прогнозированию солнечных вспышек и оценки компоненты Bz в КВМ. В настоящее время СОЛСИТ находится в стадии строительства и его эффективность для целей прогнозирования КП не подтверждена.
5. Вспомогательные наблюдательные пункты в России числом до 5 наблюдений, оснащенные автоматическим солнечным оптическим патрульным телескопом в линиях H-альфа и CaIIK.
6. Телескопы "классической" Службы Солнца: фотогелиограф (Уссурийск, Кисловодск, КрАО); фильтровый телескопы в линии H-альфа (Иркутск, Кисловодск, КрАО); измерения магнитных полей солнечных пятен магнитометрическим методом (КрАО). Наблюдения солнечной короны в линиях 5303 и 6374А (Кисловодск). Радиотелескоп РАТАН-600, для прогноза солнечных вспышек. Эти телескопы и виды наблюдения предназначены для продолжения долговременных рядов солнечной активности, картирования солнечной активности, прогноза экстремальных событий и оценки общего состояния.

**4. Модели для прогнозирования КП и центр прогноза**.
Наблюдательная сеть Службы Солнца должна быть частью общей Системы геофизического мониторинга (СМГФО). Для получения информации требуется организация наблюдений, которые охватывают очень широкий спектр электромагнитных и корпускулярных излучений – от гамма и рентгеновских лучей до радиоволн, от ядерной и электронной компонент солнечных космических лучей (СКЛ) до высокоэнергичных частиц галактических источников. Естественно, что ряд наблюдений может быть выполнен только с использованием космических аппаратов (КА): это космический сегмент наблюдательной сети СМГФО. Именно он должен обеспечивать получение непосредственной информации о тех видах солнечного излучения, которые поглощаются атмосферой и не доходят до уровня земли.

Наземный комплекс подсистемы мониторинга солнечной активности наиболее эффективен для наблюдений в оптическом и радиодиапазоне, когда предъявляются высокие требования к чувствительности, пространственному (угловому) и спектральному (частотному) разрешению. Средства мониторинга солнечной активности должны обеспечивать получение в заданном режиме наблюдательных данных для оценки и прогноза геоэффективных последствий деятельности Солнца. Они включают информацию о процессах на разных уровнях в атмосфере Солнца – фотосфере, хромосфере, короне, а также в солнечном ветре, как непосредственном продолжении солнечной короны. Схема системы прогнозирования параметров прогнозирования КП на основе национальных средств наблюдений представлена на Рис.12.

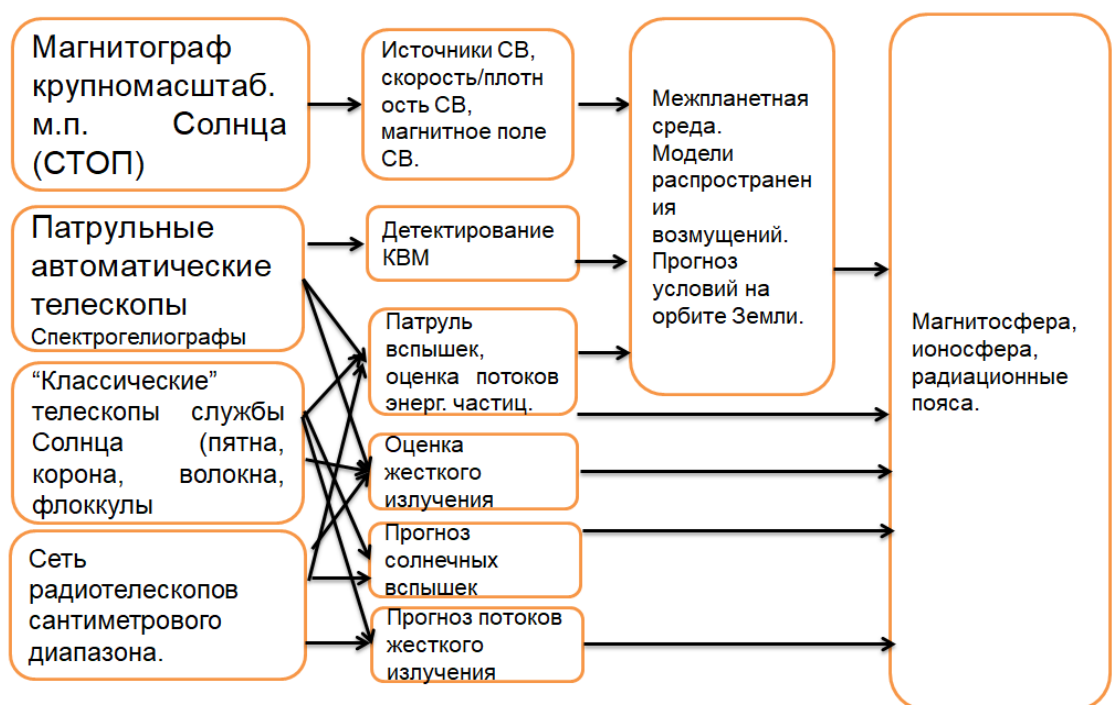

**Рис. 12**. Схема прогнозирования КП на основе наземных наблюдений.

Для прогнозирования параметров КП необходимо создание центра сбора данных, выполнять анализ данных наблюдений, проводить моделирование и представлять результаты прогнозов для пользователей и заинтересованных организаций. Центр моделирования КП осуществляет координацию работ по разработке перспективных моделей. В том числе модели: прохождения КВМ в гелиосфере; расчета параметров солнечного ветра и геоэффективности солнечных событий на основе упрощенных или полномасштабных магнитогидродинамических моделей; определения параметров рекуррентных потоков высокоскоростного солнечного ветра; вероятности развития солнечных вспышек и КВМ, по данным оперативных наблюдений и др. видов гелио и геофизических данных.

На первом этапе необходимы модели анализа данных наблюдений магнитографов и патрульных телескопов, в том числе:

1) получение информации о магнитных полях по данным наблюдений магнитографов. Методы определения интенсивности магнитного поля на основе методов спектрополяриметрии. Программы сборки синоптических карт магнитного поля, методика коррекции интенсивности магнитного поля в полярных областях Солнца.
2) Модели реконструкции корональных магнитных полей в потенциальном и не потенциальном приближениях.
3) Модель оценки скорости солнечного ветра вблизи Солнца WSA или аналогичные ей.
4) Модели распространения солнечного ветра в гелиосфере.
5) Модели детектирования эруптивных процессов и КВМ и определение параметров КВМ на начальном этапе разлета.
6) Модели распространения КВМ в гелиосфере.
7) Модели оценки геоэффективности КВМ и высокоскоростных потов СВ при достижении ими окрестности Земли.

Заметим, что модели и методы 1-6 в той или иной мере реализованы на ГАС ГАО.

Центр сбора данных и прогнозирования должен работать в непрерывном режиме. Моделирование рекуррентных потоков солнечного ветра должны осуществляться по мере поступления новых данных с магнитографов, то есть несколько раз в день. Данные о регистрации КВМ и определение их характеристик на начальном этапе их удаления от Солнца должны вноситься в модель сразу по получение данных наблюдений.

## 5. Этапы реализации проекта.

Для реализации проекта целесообразно принять план действий, который можно разбить на этапы. Первый этап длительностью 2 года может включать:

1. Создать сеть наблюдений магнитных полей Солнца. Для этого необходимо завершить введение в эксплуатацию магнитографов СТОВ в Уссурийске и Иркутске.

2. Изготовить и установить в Уссурийске, Иркутске, Крыму автоматические патрульные телескопы-спектрогелиографы. Модернизировать патрульный телескоп в Кисловодске.

3. Изготовить и установить в Уссурийске, Иркутске, КрАО, Кисловодске автоматические радиотелескопы в сантиметровом диапазоне.

4. Отладить взаимодействие наблюдательной сети. Создание методик и моделей для прогнозирования параметров КП.

Второй этап длительностью 3 года может включать:

1. Изготовление и установка в вспомогательных наблюдательных пунктов на территории России патрульных оптических телескопов.

2. Оборудование наблюдательного пункта в западном полушарии оснащенного патрульным оптическим и радиотелескопом

## 6. Наземная наблюдательная сеть для прогнозирования КП в США.
### 6.1. Существующие и перспективные наблюдательные сети.

Synoptic observations could be taken from a network of standardized telescopes whose longitudinal locations enables continuous (24 hours) duty cycle. Several solar networks are currently in operation. Solar Observing Optical Network (SOON) includes three stations situated in USA (Holloman Airforce Base, near Alamogordo, New Mexico), Australia (Royal Australian Air Force base, Learmonth, Western Australia), and San Vito dei Normanni Air Station (Italy). Routine observations include identification of solar active regions, measurement of their areas, and imaging in H$\alpha$ spectral line. Historical data sets of solar active region measurements from SOON stations can be found at NOAA National Center for Environmental Information (NCEI) at https://www.ngdc.noaa.gov/stp/solar/solardataservices.html and H$\alpha$ observations can be access via https://www.soonar.org. Monitoring of radio flux from the Sun is done using Radio Solar Telescope Network (RSTN) with four instruments located in USA (Sagamore Hill, Massachusetts and Palehua, Hawai'i), Western Australia (Learmonth), and Italy (San Vito). More information about RSTN including radio-frequency coverage see https://www.ngdc.noaa.gov/stp/space-weather/solar-data/solar-features/solar-radio/rstn-1-second/documentation/rstn-seon-usaf\_ktegnell\_31mar16.pdf. Together, SOON and RSTN facilities form Solar Electro-Optical Network, or SEON [Fitts & Loftin1993].

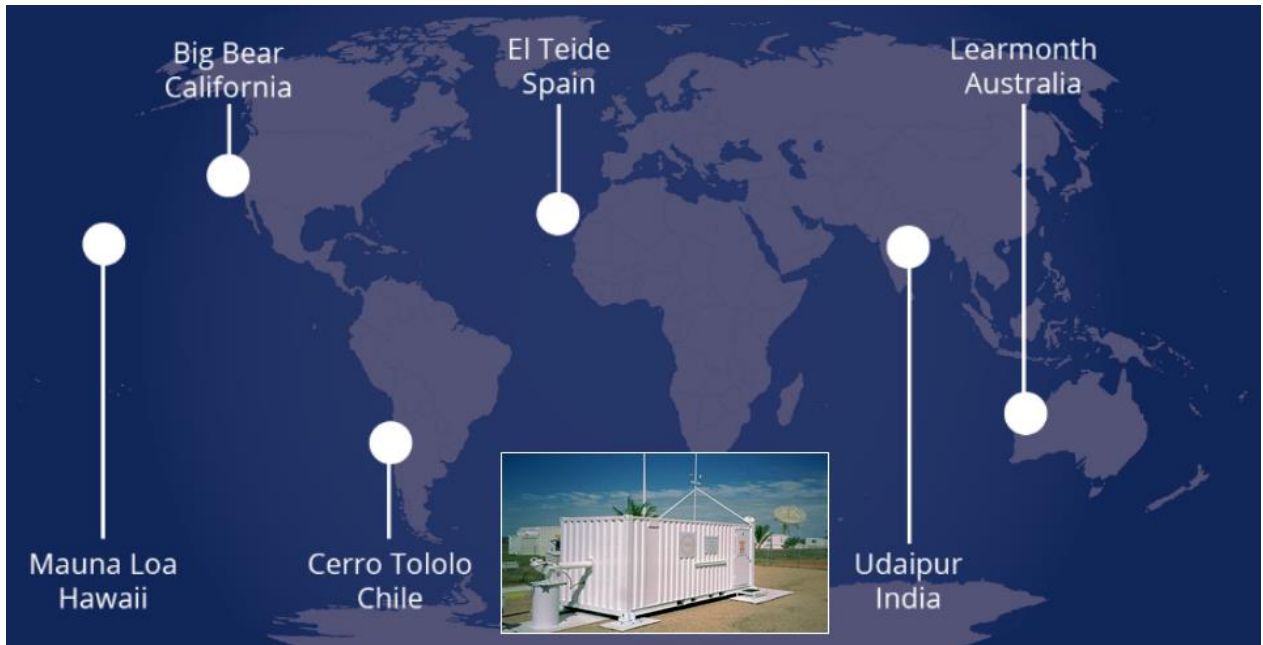

**Fig. 13**. 6-site GONG network with one of the nodes shown as inset. Background image is courtesy of Dr. C. Raftery.

Global Oscillations Network Group (GONG,Fig. 13. is a 6-site network with instruments located in USA (Big Bear Solar Observatory, California and Mauna Loa Observatory, Hawai'i), Western Australia (Learmonth), India (Udaipur Solar Observatory), Spain (Observatorio del Teide, Canary Islands), and Chile (Cerro Tololo Interamerican Observatory). Each GONG station is built on the basis of standard shipping container with all instruments situated in the optical table inside GONG enclosure. External light feed consists of two flat mirrors tracking the Sun in elevation and cross-elevation axes that feed light horizontally into the container. The optical system of the light feed is sealed by a filtered window and has an effective aperture of 2.8~cm. GONG observations are used for studies of solar interior using methods of helioseismology, evolution of magnetic fields in solar photosphere, and solar eruptive events. In addition, GONG magnetograms and H$\alpha$ images are used for the operational space weather forecast by the NOAA Space Weather Prediction Center (SWPC), the US Air Force 557$^{th}$ Weather Wing, UK Met Office, and Japan's Space Weather Canter operated by the National Institute of Information and Communications Technology (NICT). Figure 14 provides example of GONG observations and their use for operational space weather forecast at SWPC. Observations of Doppler velocities and longitudinal magnetograms are taken in the Ni I 676.8 nm spectral line at a 60 second cadence. In addition, full disk images of solar chromosphere are taken every 60 seconds in the core of H$\alpha$. Observation times for the adjacent sites are offset by 20 seconds, and thus, the network cadence for H$\alpha$ images is 20 seconds. Each node automatically powers-on shortly after sunrise, finds and locks on solar disk for a full day of observations, and it shuts down at the sunset. Each node is equipped with weather station and solar irradiance monitor, which are used in a decision tree for starting/ending daily observations assisted by the local support staff, if/when needed. Each station is equipped by GPS unit that provides a high-precision time, which is necessary for interpretation of helioseismic network observations. GONG network is operated by the US National Solar Observatory's integrated Synoptic Program (NISP). Based on 18 years of observations, 6-site GONG network achieves the mean duty cycle 88.3% and the median value is 92.0%. Furthermore, the data show no long-term variations in duty cycles, which implies that there appears no significant changes in climate in any of the GONG network sites (Jain.etal2021). For additional information about GONG see [Pevtsov 2016, Hill 2018].

Global High-Resolution H-alpha Network (Steinegger.etal2000) is *ad hoc* network of H$\alpha$ instruments operated by different observatories. The telescopes may have different parameters (e.g., aperture, filter bandpass, cadence, data reduction etc), and take independent observations. The reduced images are sent to a common database.

Several networks are in different stages of development, including the Solar Activity Monitor Network (Erdelyi.etal2022), Continuous Halpha Imaging Network [Ueno.etal2010], the Solar Physics Research Integrated Network Group [Gosain.etal2018], and next generation GONG (ngGONG). SamNet is a global network of small aperture telescopes equipped with magneto-optical filters (MOF). The main emphasis of SamNet is a flare prediction based on parameters derived from photospheric magnetograms of longitudinal magnetic field. CHAIN concentrates on full disk observations of solar chromosphere in H-alpha with emphasis on solar eruptive events (flares, erupting filaments, and flare-related Moreton waves). As part of the network development, the Flare Monitoring Telescope (FMT), which was in operations since 1992 at Hida Observatory (Japan) was relocated in 2010 to Ica University (Peru). FMT is a ``battery'' of identical refractor telescopes installed on a single telescope mount. Each telescope has an effective aperture of 64 mm. The observations are taken with 30 sec cadence and use Fabry-P'erot etalons tuned at red continuum, H$\alpha$ core 656.3 nm, and H$\alpha$ wings ($\pm 0.08$ nm). In 2017, a second FMT instrument was installed in King Saud University (Saudi Arabia).

Together with the Solar Magnetic Activity Research Telescope (SMART) at Hida Observatory (Japan), these three instruments form current CHAIN network. There are plan to install third FMT instrument in Algeria..

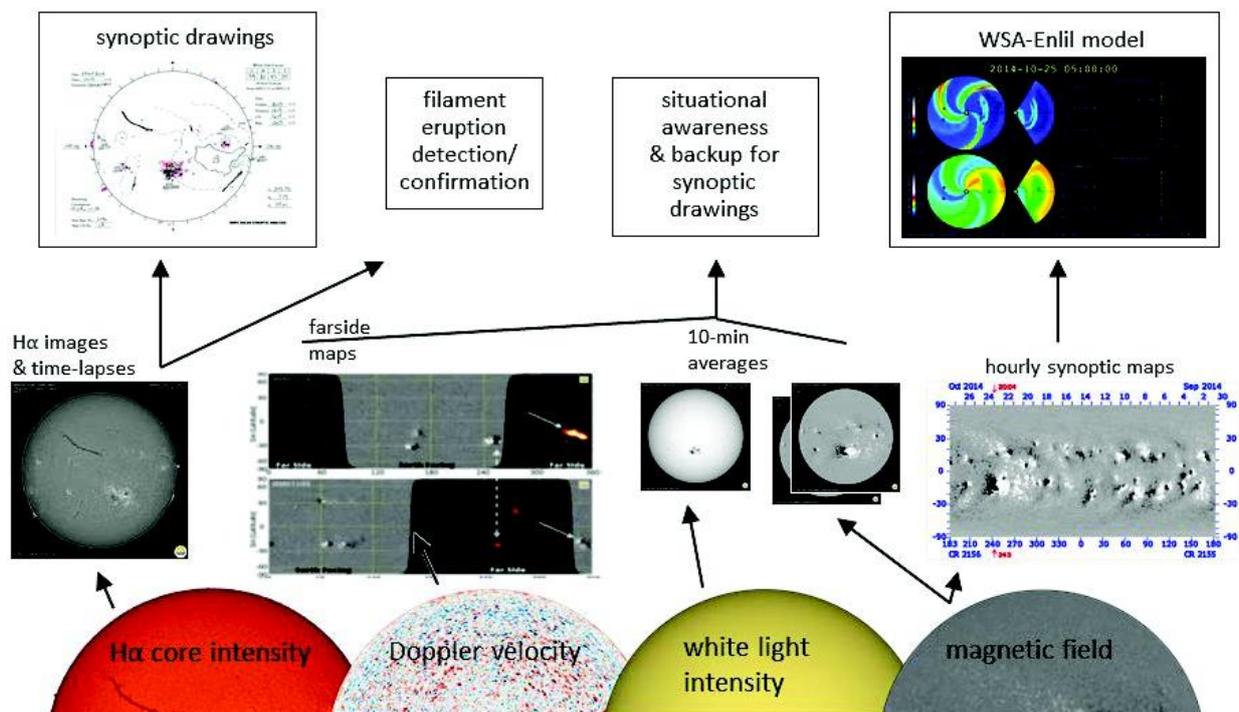

**Fig. 14**. Example of GONG observations (bottom row), the derived data products (middle row), and their application to operational space weather forecast and modeling (upper row). GONG takes full disk observations, but for this representation, only half of full disk images are shown. Courtesy of Dr. A. Marble

List of groundbased networks summarized in this Section is not inclusive. For example, it does not include the Birmingham Solar-Oscillations Network [Hale.et al. 2016], and its future development [Hale.et al. 2020]. International Network of Solar Radio Spectrometers called Compound Astronomical Low frequency Low cost Instrument for Spectroscopy and Transportable Observatory [Benz et al. 2009]. For a recent review of radio-wavelength facilities dedicated to space weather see [Carley et al. 2020]. Geographic location of all solar synoptic facilities currently in operations could be found at https://goo.gl/LRVhVk.

## 6.2. Next generation GONG
6.2.1. Need for a next generation network GONG network

GONG is aging, 26+ years in operations GONG refurbishment will extend lifetime by 5-7 years new cameras, improved polarization modulators, Data Center upgrades, new workstations, shelter cooling system upgrades, misc. New science and model requirements: Vector magnetic field in photosphere and chromosphere Doppler velocity for helioseismology (higher resolution, multi-height) Polarimetry in the corona Sun-as-a-star+disk-resolved spectra (same spectral lines) Full disk (high duty cycle) and long-duration (1-2 full magnetic cycles).

## 7. Обсуждение.

В данной работе представлена схема организации и комплекс мер для обеспечения прогнозирования космической погоды в нашей стране. На данном этапе основную роль могут играть наземные наблюдения, на основе долготно распределенной сети наблюдательных пунктов. Существует два основных вида наблюдений, которые должны выполняться. Это наблюдения крупномасштабных магнитных полей и регистрация КВМ и солнечных вспышек.

Магнитографы крупномасштабного поля созданы и установлены в Уссурийске, Иркутске и Кисловодске. Регистрация КВМ может осуществляться патрульными спектрогелиографами. Прототипы таких телескопов созданы и эксплуатируются на ГАС ГАО в Кисловодске.

Следует отметить, что состав наблюдательных данных предлагаемой службы прогнозирования КП близок к схеме принятый в США в SWPC/NOAA, а также перспективным видам наблюдательных программ в США. Основные отличия в методе регистрации КВМ. В SWPC это орбитальные коронографы, в нашей схеме наземные спектрогелиографы.

Помимо этих видов наблюдений могут и должны осуществляться и другие виды наблюдения. Прежде всего, это виды наблюдений, продолжающие "классические" ряды Службы Солнца. Это наблюдения фотосферы и подсчет характеристик солнечных пятен (индекс и площадь солнечных пятен), наблюдения солнечной короны, наблюдения магнитных полей солнечных пятен, продолжение рядов характеристик солнечных волокон, протуберанцев, флоккул в линии CaIIK. Эти виды оптических наблюдений должны быть стандартизированы, что важно для сохранения и стабильности долговременных рядов солнечной активности. В годы советской Службы Солнца данные различных обсерваторий взаимно дополнялись. Издавался единый бюллетень "Солнечные данные". В нынешних условиях это может общий Интернет сайт. Прототип такого сайта по данным ежедневных наблюдений ГАС создан (https://observethesun.com/). Необходимо, чтобы данные сайта дополнялись данными обработки наблюдений других солнечных обсерваторий.

Существуют перспективные виды наблюдений, которые могут существенно расширить улучшить прогноз солнечной активности для целей КП. Это телескоп-магнитограф полного вектора магнитного поля с высоким пространственным разрешением СОЛСИТ/ИСЗФ, радиогелиограф ССРТ/ИСЗФ, РАТАН-600/САО. Благодаря этим видам наблюдений будет возможность осуществлять наблюдения и прогноз солнечных вспышек. Важным видом наблюдений являются наблюдения в сантиметровом радиодиапазоне малым радиотелескопами. Сеть радиотелескопов позволит фиксировать солнечные вспышки даже в плохие погодные условия.

В случае создания наблюдательной сети для обеспечения прогнозирования КП будет накоплен и предоставлен исследователям для анализа большой объем данных. Уже сейчас объем данных, накопленный на ГАС составляет более 400 Тбт. Поскольку сеть будет работать в непрерывном режиме, это позволит отслеживать и проводить анализ большинства быстропротекающих и высокоэнергичных событий, таких как эрупцию вещества, предвспышечных явлений, изучать развитие активных областей и другие процессы. Также патрульные наблюдения позволяют осуществлять анализ локальных глобальных осцилляций, а следовательно выполнять исследования по фундаментальным задачам астрофизики. Наблюдения поля скорости на полном диске Солнца с характерным временем накопления 1-2 минуты на сегодняшний момент являются уникальными. Таким образом, данные наблюдений новой наблюдательной сети могут существенно расширить общемировые ресурсы данных наблюдений о солнечной активности.

Новые виды наблюдений потребует создание нового программного обеспечения по обработке данных, оперативному моделированию и прогнозированию, а также Интернет ресурса для представления данных о солнечной активности и космической погоде. Это будет способствовать общему развитию солнечной физики и астрофизики в нашей стране.

**Литература**

**Приложение. 1**

Расположение наблюдательных пунктов и их оснащение приведены и наблюдательные программы в Таблице 1.

**Табл. 1**. Возможные наблюдательные пункты "Службы Солнца" с указанием их статуса (основные и другие); долгота; перечень необходимых телескопов; перечень дополнительных средств наблюдений[*]

| N/тип | Название/долгота | Оснащение по сл. Солнца/ Текущие наблюдения | Существ. телескопы |
|---|---|---|---|
| 1 осн. | Уссурийская астрофизическая обсерватория ДВО РАН (L=132°) | 1) магнитограф (уст.) <br> 2) авт. пат. телескоп. <br> 3) авт. рад. Телескоп <br><br> Наблюдения: <br> Солнечные пятна 1954-2020 <br> Радиотелескоп РТ-2 1990-0.6.2022 <br> Магнитные поля пятен 1966-1989 <br> С июля 2022 г. Наблюдений нет. <br> . | Больш. корон.(БСВТ) <br> Горизонтальный Солнечный Телескоп <br> Радиотелескоп РТ-2 <br> Радиотелескоп РТ-8 <br> Хромосферно-фотосферный телескоп |
| 2 осн. | Иркутск, ИСЗФ, (Листвянка, L=104.8°) | 1) магнитограф (уст.) <br> 2) авт. патр. тел. <br> 3) авт. рад. тел. <br> Наблюдения в линии CaIIK. 2006, 2009 – 2021 <br> Магнитограммы 1997, 1999-2013 <br> Радионаблюдения ССРТ по настоящее время. | БСВТ <br> Нα фильт. телескоп <br> KCaII фильт. тел. <br> Вект.-магн. СОЛСИТ <br> Гориз.солн. тел. <br> СТОП-1 <br> Больш. коронограф <br> Радиотелескоп (ССРТ) |
| 3 осн. | Кисловодск, ГАС (L=42.3°) | 1) магнитограф (уст.) <br> 2) авт. патр. тел. (уст.) <br> 3) авт. рад. тел. <br><br> Все виды наблюдений | 1 Фотогелиограф <br> 2.Коронограф. <br> 3.Hα тел."Opton" <br> 4.Баш.спектрогел.KCaII <br> 5.Пат.тел.(СПОТ)KCaII. |

| | | | 7.Радиотел.4.9 и 2.9 см. |
|---|---|---|---|
| | | продолжаются по наст. время. | 8.Магнитометр |
| | | | 9.Больш. корон. БСВТ |
| | | | 10.Пат.тел. Н-альфа. |
| 4 | КрАО (L=34°) | 1) авт. пат. телескоп. 2) авт. рад. телескоп. Наблюдения в линии He10830 1999-н.в. Магнитные поля пятен 1957-н.в. Коронограф - 1 (КГ-1) Halpha 1999-2018 Коронограф - 2 (КГ-2) | БСТ-1; спектр. м.поля. БСТ-2; 10830; м.п.пят. Тел. КГ-1 Hα БСВТ Радио.тел. 10,7 см. |
| 5 | ИЗМИРАН (L=37.3°) | | 1.Радиометр на частотах (169, 204 и 3000 МГц) 2. спектрограф (25-270 МГц). |
| 5 | Коуровская обсерватория УрФУ (L=59.5°) | 1) авт. пат. телескоп. 2) авт. рад. телескоп | Солн. тел. АЦУ-5 |
| 6 | Элиста КалмГУ (L=44.2°) | 1) авт. пат. телескоп. (уст.) | авт. пат. телескоп. |
| 7 | Калиниград, ИЗМИРАН (L=20.4°) | 1) авт. рад. телескоп. | |
| 8 | Куба, Гавана (L=-82.5°) | 1) авт. пат. телескоп. 2) авт. рад. телескоп. | Антены РТ-3 РТ-2; перспективный |
| 9 | Тенериф, обс. МГУ (L=-16.3°) | 1) авт. пат. телескоп. 2) авт. рад. телескоп. | перспективный |

*Точное число наблюдательных пунктов и их местоположение может варьироваться.